\documentclass[lettersize,journal]{IEEEtran}
\usepackage{amsmath,amsfonts}
\usepackage{algorithmic}
\usepackage{algorithm}
\usepackage{array}
\usepackage[caption=false,font=normalsize,labelfont=sf,textfont=sf]{subfig}
\usepackage{textcomp}
\usepackage{stfloats}
\usepackage{url}
\usepackage{verbatim}
\usepackage{graphicx}
\usepackage{cite}
\usepackage{color}
\DeclareMathOperator*{\argmin}{argmin}
\begin{document}

\title{Spatial Deep Deconvolution U-Net for Traffic Analyses with Distributed Acoustic Sensing}

\author{Siyuan Yuan, Martijn van den Ende, Jingxiao Liu, Hae Young Noh, Robert Clapp, Cédric Richard, Biondo Biondi
\thanks{Siyuan Yuan, Jingxiao Liu, Hae Young Noh, Robert Clapp, and Biondo Biondi are with Stanford University, CA, U.S.A.}
\thanks{Martijn van den Ende, and Cédric Richard are with the Université Côte d’Azur,, OCA, UMR Lagrange, France. }
\thanks{This work has been submitted to the IEEE for possible publication. Copyright may be transferred without notice, after which this version may no longer be accessible.}
}



\maketitle
\begin{abstract}

Distributed Acoustic Sensing (DAS) that transforms city-wide fiber-optic cables into a large-scale strain sensing array has shown the potential to revolutionize urban traffic monitoring by providing a fine-grained, scalable, and low-maintenance monitoring solution. However, the real-world application of DAS is hindered by challenges such as noise contamination and interference among closely traveling cars.  In response, we introduce a self-supervised U-Net model that can suppress background noise and compress car-induced DAS signals into high-resolution pulses through spatial deconvolution. Our work extends recent research by introducing three key advancements. Firstly, we perform a comprehensive resolution analysis of DAS-recorded traffic signals, laying a theoretical foundation for our approach. Secondly, we incorporate space-domain vehicle wavelets into our U-Net model, enabling consistent high-resolution outputs regardless of vehicle speed variations. Finally, we employ L-2 norm regularization in the loss function, enhancing our model's sensitivity to weaker signals from vehicles in remote traffic lanes.
We evaluate the effectiveness and robustness of our method through field recordings under different traffic conditions and various driving speeds. Our results show that our method can enhance the spatial-temporal resolution and better resolve closely traveling cars. The spatial deconvolution U-Net model also enables the characterization of large-size vehicles to identify axle numbers and estimate the vehicle length. Monitoring large-size vehicles also benefits imaging deep earth by leveraging the surface waves induced by the dynamic vehicle-road interaction.
\end{abstract}



\section{Introduction}
\IEEEPARstart{T}RAFFIC monitoring systems, which automatically and continuously detect, track, and characterize vehicles in moving traffic, provide valuable information for urban management, maintenance, and planning. Conventional monitoring systems include vision-based \cite{5309837, 7458203,REINARTZ2006149} and pavement sensing technologies (e.g., inductive loops~\cite{jain2019review,6957957,doi:10.3141/1719-14} and piezoelectric sensors~\cite{zhang2015new,li2006application,jain2019review}). These approaches are well-developed but have several drawbacks. For example, camera systems bring individual-privacy concerns and are sensitive to weather conditions; point pavement sensing systems provide spatially sparse sampling and are challenging to maintain. An additional drawback of these point-sensing systems is their requirement for on-site installation. Recent mobile sensing methods overcome the on-site installation and maintenance challenge, but such systems require cellular connectivity and users to opt-in location tracking.

Our work addresses these problems by leveraging an emerging fiber-optic sensing technology called DAS. This technology turns optical fibers into dense-sampling seismic recording arrays, offering several benefits. First, DAS records absolute signals of human-induced deformation that cannot be easily tied to any individual. Second, DAS monitoring can be cost-efficient for city-scale monitoring by leveraging ubiquitous pre-existing telecommunication infrastructures as sensors. Finally, the system relies on a single optoelectronic device known as an interrogator to be deployed in a secure and easily accessible location. The DAS approach is in stark contrast to the conventional methods having numerous individually powered instruments scattered across a city, exposed to meteorological conditions and detrimental interactions with humans, plants, and animals.

\subsection{Background on DAS}

DAS-based monitoring is performed by connecting an interrogator to one end of a standard telecommunications‐grade optical fiber. The interrogator sends short laser pulses into the optical fiber and measures the subtle phase shifts of Rayleigh scattered light returning to the detector at a predicted two‐way travel time~\cite{Posey, MasoudiAndNewson}. In this way, the strain field induced by natural processes (e.g. earthquakes) and urban activities (e.g. moving vehicles, construction, pumps) acting on the fiber coupled to the Earth can be sampled at a meter‐scale spatial resolution over tens of linear fiber kilometers. Notably, DAS technology has proven effective on the pre-existing ``dark fiber" telecommunication infrastructure for earthquake monitoring \cite{biondi2021}, infrastructure monitoring \cite{liu2023bridge, YuanEtAL2021}, and near-surface imaging \cite{EMartin2017, Fang2020}, significantly reducing installation cost. Furthermore, by incorporating state-of-the-art interrogators, such as OptaSense QuantX\cite{quantx}, DAS can record full waveform signals over distances up to 50 km, with a granular 1 m channel spacing, which makes DAS promising for continuous city-wide sensing \cite{biondi2021b}. 



 Despite the aforementioned advantages, analyzing data from a DAS array located in an urban environment has been challenging because it records a complex mixture of inherently unlabeled signals \cite{Huot2018}. As such, researchers have been developing algorithms that try to detect vehicles accurately and automatically from DAS recordings.

\subsection{Existing vehicle tracking and detection algorithms}
\cite{LindseyEtAl2020} applied a common seismological method, the short-time-average through long-time-average trigger (STA/LTA). To exploit the array geometry of DAS, beamforming algorithms have been applied to detect cars and measure their speed \cite{Wang2021, VanDenEnde2021b}.  These simple methods performed well on roads with relatively light traffic and without complicated traffic patterns. \cite{WiesmeyrEtAl} took this a step further by utilizing the Hough transform, a technique derived from the field of image processing, to estimate the flow and average speed of vehicles. \cite{m2pi2020} introduced data mining and signal processing methods, such as clustering algorithms and Kalman filtering techniques, to identify and track vehicles.  A novel contribution by \cite{Yeetal} was the application of a real-time object detection algorithm based on deep learning techniques for estimating traffic flow and vehicle speed from DAS data, which was assessed along a 500-meter fiber length in the suburbs of Beijing. More recently, \cite{LiuEtAl} develop a spatial-domain Bayesian filtering and smoothing algorithm to detect, track, and characterize each vehicle to obtain fine-grained traffic monitoring with DAS. Despite their limited success, these methods gradually become inaccurate if many vehicles transit simultaneously close to the same segment of the fiber cable, as signals from different cars start overlapping.

\subsection{Recent advancement that improves the resolution of vehicle-induced signals}
A parallel research direction from \cite{van2021deep} aims to improve the resolution of vehicle signals recorded by DAS by reducing interference among closely traveling cars. The research is valuable as a preprocessing step for algorithms to further improve their detection and tracking ability. Specifically, \cite{van2021deep} proposed a self-supervised time-domain deconvolution Auto-Encoder (time-domain DAE) with the vehicles' impulse responses from the quasi-static recordings. Compared to a conventional channel-wise deconvolution algorithm, the time-domain DAE model has the benefit of incorporating the spatial-temporal characteristics of car signals, leading to much sharper and more localized outputs. The authors applied a beamforming algorithm to the localized outputs rather than the original inputs. They showed significant improvements in terms of the resolution in car-speed estimation and detection accuracy. Furthermore, as shown in \cite{van2021deep}, the time-domain DAE model, once trained, can deconvolve 24-hour recordings in less than 30 seconds, achieving $>$ 400 times speedup compared to a conventional iterative approach, which makes the method promising for real-time processing.

However, the time-domain DAE model assumes a stationary Ricker wavelet in the time domain as the vehicle's impulse response. This assumption can undermine vehicle-tracking accuracy. As the following sections of this paper will show, vehicles' temporal wavelets are non-stationary to speed variation. Due to the wavelet mismatch, a time-domain DAE model that works well with fast-speed traffic produces low-resolution results for low-speed vehicles.

\subsection{Novelties of our work}
This study overcomes the constraints of \cite{van2021deep} by developing a new algorithm that delivers precise results for vehicles traveling at differing speeds. The study's contribution is threefold.

Our first contribution is a comprehensive analysis of the resolution of DAS-recorded traffic signals, based on both physics and field studies. This key element forms the theoretical bedrock of our method, with a detailed discussion provided in Section II.

Secondly, we introduce a novel departure from the methodology in \cite{van2021deep}. Contrary to the use of a Ricker vehicle wavelet that remains stationary in the time domain, we integrate a physics-based vehicle wavelet in the spatial domain into our space-domain DAE model. A significant benefit of employing a spatial wavelet is that it enables our model to produce uniformly high-resolution outputs, irrespective of changes in the speed of the vehicle. A detailed discussion is provided in the Methods Section.

The third unique aspect of our study involves the optimization of the regularization term in the loss function. \cite{van2021deep} used the L-1 norm to enhance the sharpness of the outputs. Similarly, we employed the L-1 norm to amplify the output sharpness. However, we noticed that the L-1 norm might miss out on the weak energy released by vehicles traveling on far-off traffic lanes. Consequently, we investigated the use of the L-2 norm regularization, which enabled us to establish optimal settings to identify weaker signals from vehicles on more distant traffic lanes.



The rest of the paper is organized as follows. Section II demonstrates the resolution reduction of traffic-induced DAS signals due to a combined effect of physics and the limitation of the sensors. Through the numerical characterization of vehicle impulse responses and observations from field experiments, we design a spatial DAE model aiming to improve the signal resolution for speed-varying vehicles. Section III describes our spatial DAE model in detail. Section IV describes the training procedure and baseline approaches. Section V benchmarks the proposed DAE model with baselines for car motion and speed tracking. Section VI concludes our work and describes future direction.

\section{Vehicle-Induced DAS Response}
This section investigates the vehicle-induced DAS response to inform the design of our proposed space-domain DAE model. We first review two categories of vehicle-induced DAS responses, quasi-static and surface-wave signals, as a basis for our study. Subsequently, we carry out a theoretical analysis and numerical simulations which reveal a resolution decline in traffic signals due to physical constraints and sensor limitations. This discovery drives the need for resolution enhancement through our novel deconvolution algorithm. Finally, we recognize the speed-invariance of the spatial wavelet, a pivotal observation that motivates our development of the space-domain DAE model. By incorporating the spatial wavelet into our model, we are able to achieve better accuracy with speed variation and better resolution for vehicle tracking. 
\subsection{Quasi-static v.s. dynamic response}
Permanently deployed fiber-optic cables are often buried (trenched) or placed within underground conduits. We assume here that the DAS fiber is deployed alongside a road at some depth below the surface. When a vehicle passes near-by the virtual sensors of the roadside telecom fiber cable, the interaction between the vehicle and the road structure induces the deformation of the telecom fiber cable. The signal pattern of vehicle-induced telecom fiber deformation is a function of the vehicle characteristics, fiber conduit properties, ambient conditions, etc. There are mainly two components of signals produced by moving vehicles: quasi-static signals ($<$ 1 Hz) resulting from the ground deformation due to the vehicle's weight, and propagating surface waves (2 to 20 Hz) caused by the dynamic vehicle-road interaction resulting from the roughness of the road (e.g., bumps).

Previous studies~\cite{jousset2018dynamic, LindseyEtAl2020, YuanEtAl2020} have found that the quasi-static component dominates the energy of vehicle-induced telecom fiber vibration and is theoretically described by the Flamant-Boussinesq approximation \cite{Fung}. As a vehicle approaches the virtual sensor, ground deformation above the sensor increases, and the fiber coupled to the earth is stretched, resulting in increased tension in the fiber. As the vehicle moves away, ground deformation near the virtual sensor and the fiber tension decreases. Due to the relatively strong energy, simplicity, and compactness compared to the surface-wave component, quasi-static signals have been used for car tracking and detection tasks \cite{van2021deep}. For the same reason, our car tracking method is also based on the quasi-static response.  

\subsection{Sensing resolution limited by physics and sensor limitation}
\begin{figure}[t]
    \centering
    \includegraphics[width=1\linewidth]{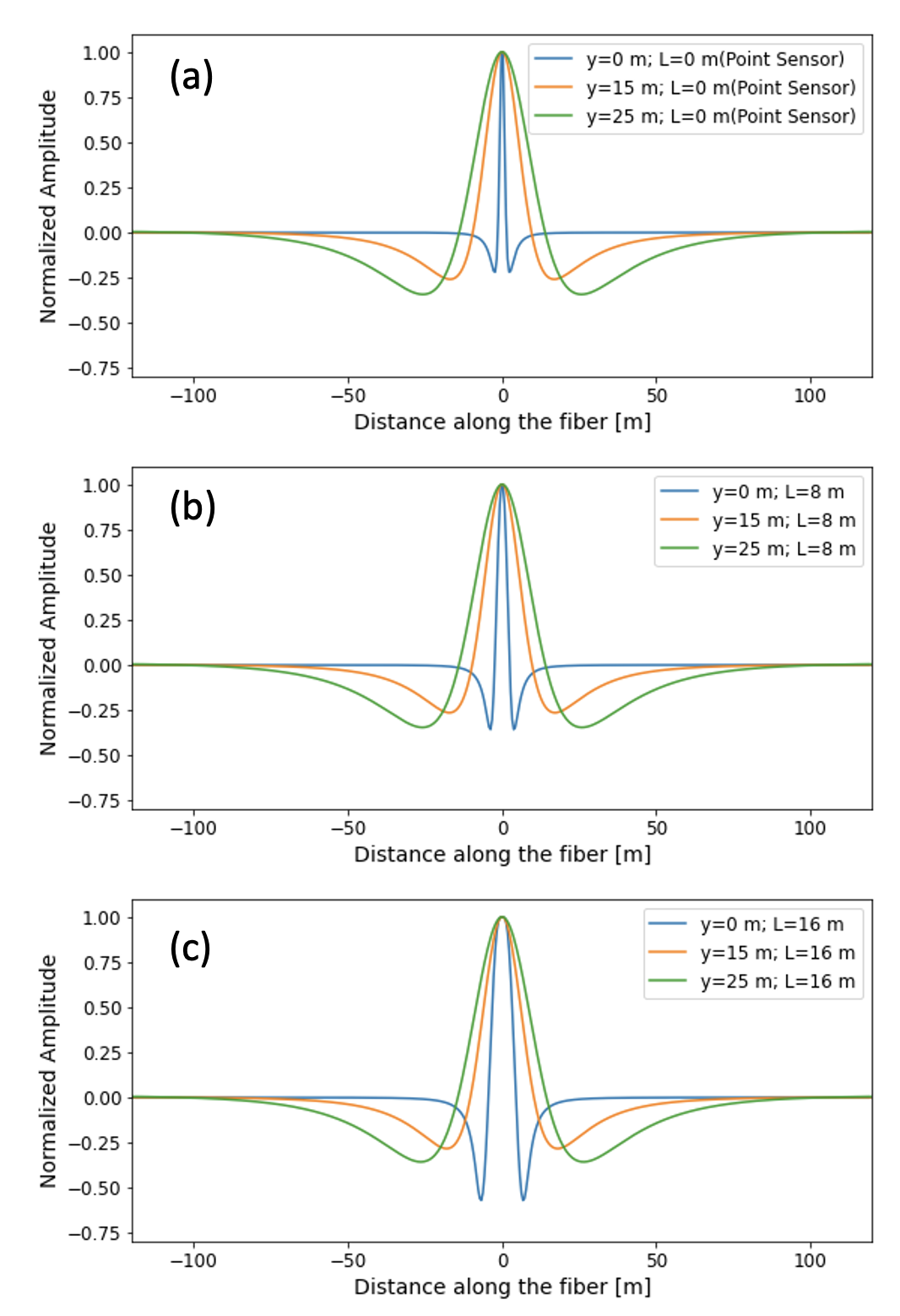}
    \caption{Simulation of quasi-static signals of a DAS array subjected to a point load with various lateral offsets, $y$, to the fiber. (a), (b), and (c) shows the simulated wavelets for gauge lengths, $L$, of 0, 8, and 16 m, respectively.}
    \label{fig:point_sim}
\end{figure}

\begin{figure}[t]
    \centering
    \includegraphics[width=1\linewidth]{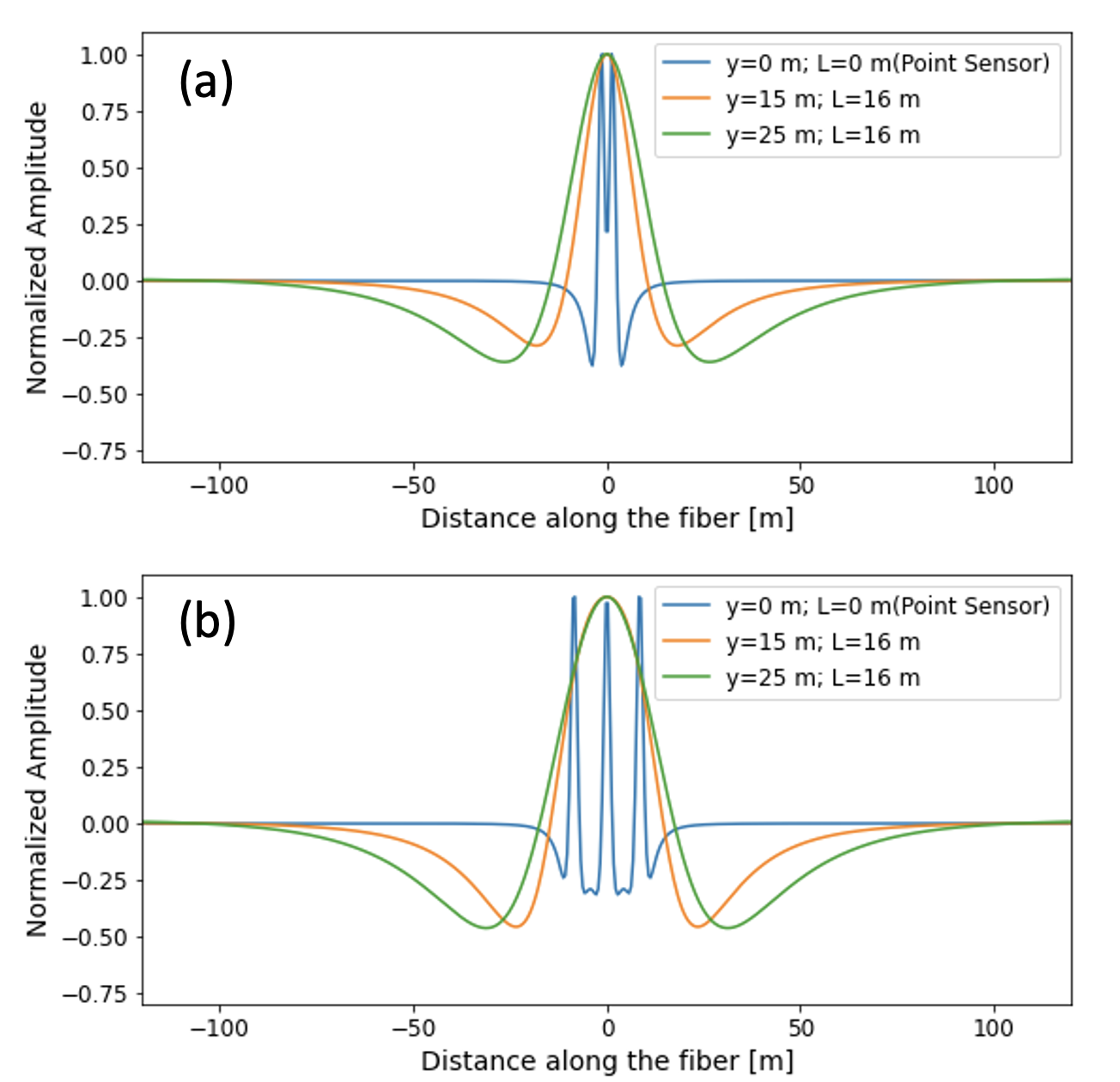}
    \caption{Simulation of quasi-static signals of a DAS array with zero and 16 m gauge length, $L$, loaded with (a) a two-axle car with an axle spacing of 2.8 m; (b) a three-axle vehicle with an axle spacing of 9 m with various lateral offsets, $y$, to the fiber.}
    \label{fig:mutiaxles_sim}
\end{figure}

We conduct synthetic experiments to demonstrate the resolution degradation of quasi-static signals due to physics and sensor limitations. We model vehicular seismic sources using a collection of vertical point forces located at the vehicle's wheel-road contacts \cite{jousset2018dynamic, LI2018147, YuanEtAl2020, LindseyEtAl2020}. Define $x$ as the horizontal distance along the fiber. The quasi-static or geodetic strain ($E_x$) DAS signal from a vehicle centered at $x=0$ is equal to the change in displacement over the DAS gauge length ($L$). Displacement in the direction of the fiber ($U_x$) can be modeled using the Flamant-Boussinesq equation for a point load applied to a half space with basic knowledge of the fiber and vehicle locations, the vehicle load ($F_z$), the soil shear modulus ($\mu$), and Poisson's ratio ($\nu$):

\begin{equation}
\label{ux}
U_x(x)=\frac{F_z}{4\pi\mu}(\frac{zx}{r^3}-\frac{(1-2\nu)}{r+z}(\frac{x}{r})),
\end{equation}

\begin{equation}
\label{Ex}
E_x(x)=\frac{U(x+\frac{L}{2})-U(x-\frac{L}{2})}{L},
\end{equation}

where $z$ is fiber depth, $r=\sqrt{x^2+y^2+z^2}$ is the 3-D distance from the fiber position to the car, and $y$ is the lateral fiber-road offset. From the equation, we can see that for fixed soil properties, the shape of spatial strain response depends on the gauge length and the relative position between the load and the fiber. To compute the modeled synthetic horizontal strain signal, we assumed a fiber depth of 2 m, appropriate values for sandy/clayey soils, such as a Poisson's ratio of 0.4. Fig. \ref{fig:point_sim} shows the simulated strain along the fiber for a point load at $x=0$ m and $y=0,15,25$ m. We investigate the gauge length effect by simulating wavelets with $L$ of 0 (equivalent to point sensor), 8, and 16 m in Fig. \ref{fig:point_sim} (a), (b), and (c), respectively. From the point sensor wavelets shown in Fig. \ref{fig:point_sim} (a), we can see that the wavelet is compact and sharp when the fiber is right beneath the load. As $y$ increases, we can observe an increasing smoothing effect leading to wider wavelets and lower resolution. By comparing Fig. \ref{fig:point_sim} (a) with (b) and (c), we can see that increasing gauge length leads to a wider wavelet for 0 offset, but has a limited impact on larger $y$ (15 and 25 m). Fig. \ref{fig:mutiaxles_sim}
 (a) and (b) show the simulation of a two-axle car with 2.8 m axle spacing and a three-axle vehicle with 9 m axle spacing, respectively. We compare the point senor recording to that obtained with $L=16$ m and the fiber-road offset of 15 m and 25 m. We can see that the axles are resolved from the point sensor recordings with zero lateral offset, but become indistinguishable in recording with a larger offset due to the smoothing effect. 

\subsection{Wavelet dependency on vehicle speed}
This subsection studies the effect of vehicle speed on the DAS response in the temporal and spatial domain, which is important to guide the design of the DAE model. 
For a vehicle traveling from an initial position, $x_0$,  with a constant speed of $c$, the measured strain response at fiber location $x$ is,

\begin{equation}
\label{uxt}
\varepsilon_x(x, t)=E_x(x-ct-x_0).
\end{equation}

\begin{figure}[t]
    \centering
    \includegraphics[width=0.8\linewidth]{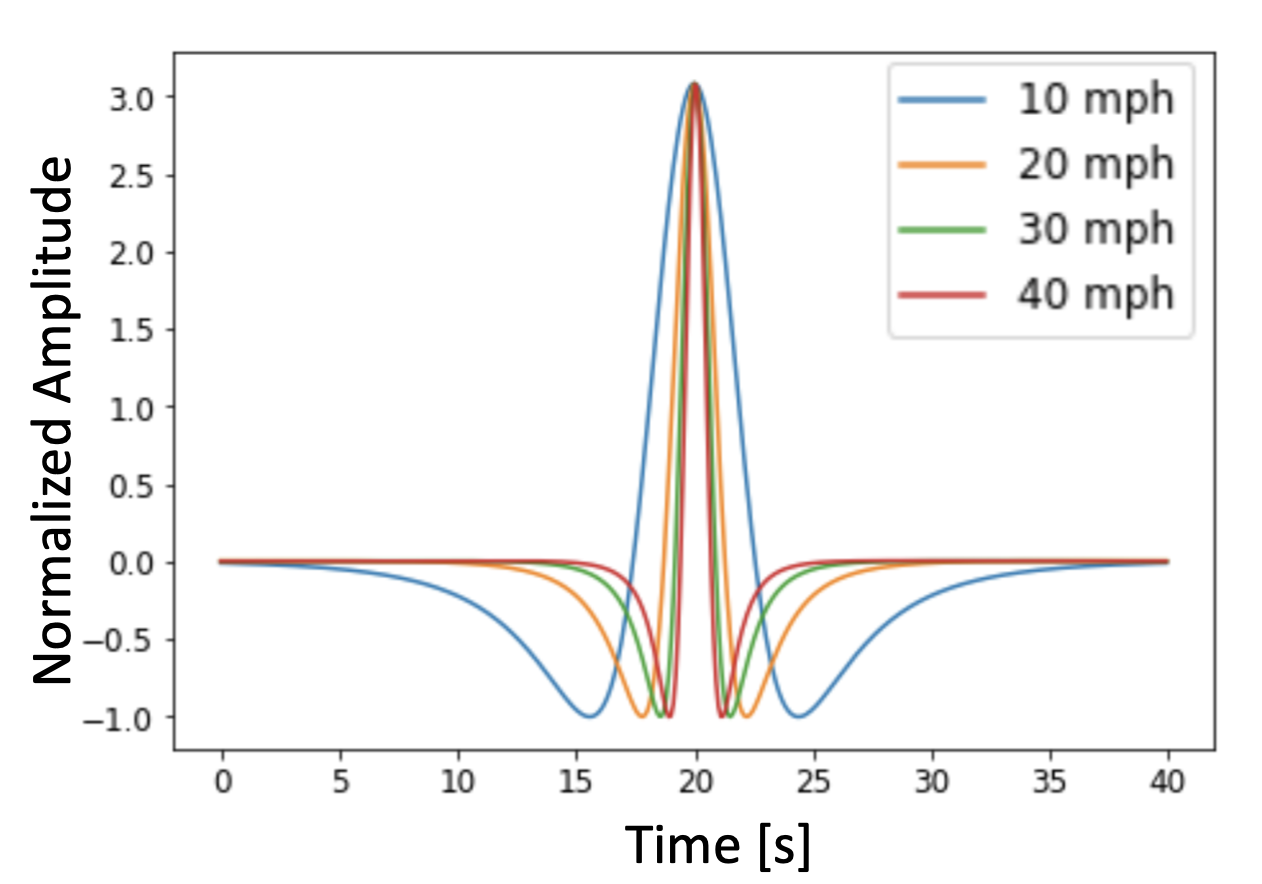}
    \caption{Numerical simulation of the time-domain quasi-static signals of a two-axle car with constant speeds of 10, 20, 30, and 40 mph.}
    \label{fig:wavelet_comparisons}
\end{figure}

The temporal wavelet at a constant time $x_0$ and the spatial wavelet at a constant location $t_0$ can be written as $\phi(t)=\varepsilon_x(x=x_0, t)$ and $\psi(x)=\varepsilon_x(x, t=t_0)$, respectively.
From equations \eqref{ux} to \eqref{uxt}, we can see that the car speed, $c$, acts as a scaling factor of $\phi(t)$, determining the compactness of the temporal wavelet. This can be confirmed through numerical experiments. Fig. \ref{fig:wavelet_comparisons} shows the simulation of the temporal wavelet of a two-axle car traveling at 10, 20, 30, and 40 mph. We can see that faster car speed leads to sharper temporal wavelet. On the other hand, we can see from the equations that the shape of the spatial wavelet is speed-independent. The speed term only introduces a space shift accounting for the different vehicle positions at time $t_0$ for different speeds.
 
\begin{figure}[t]
    \centering
    \includegraphics[width=0.8\linewidth]{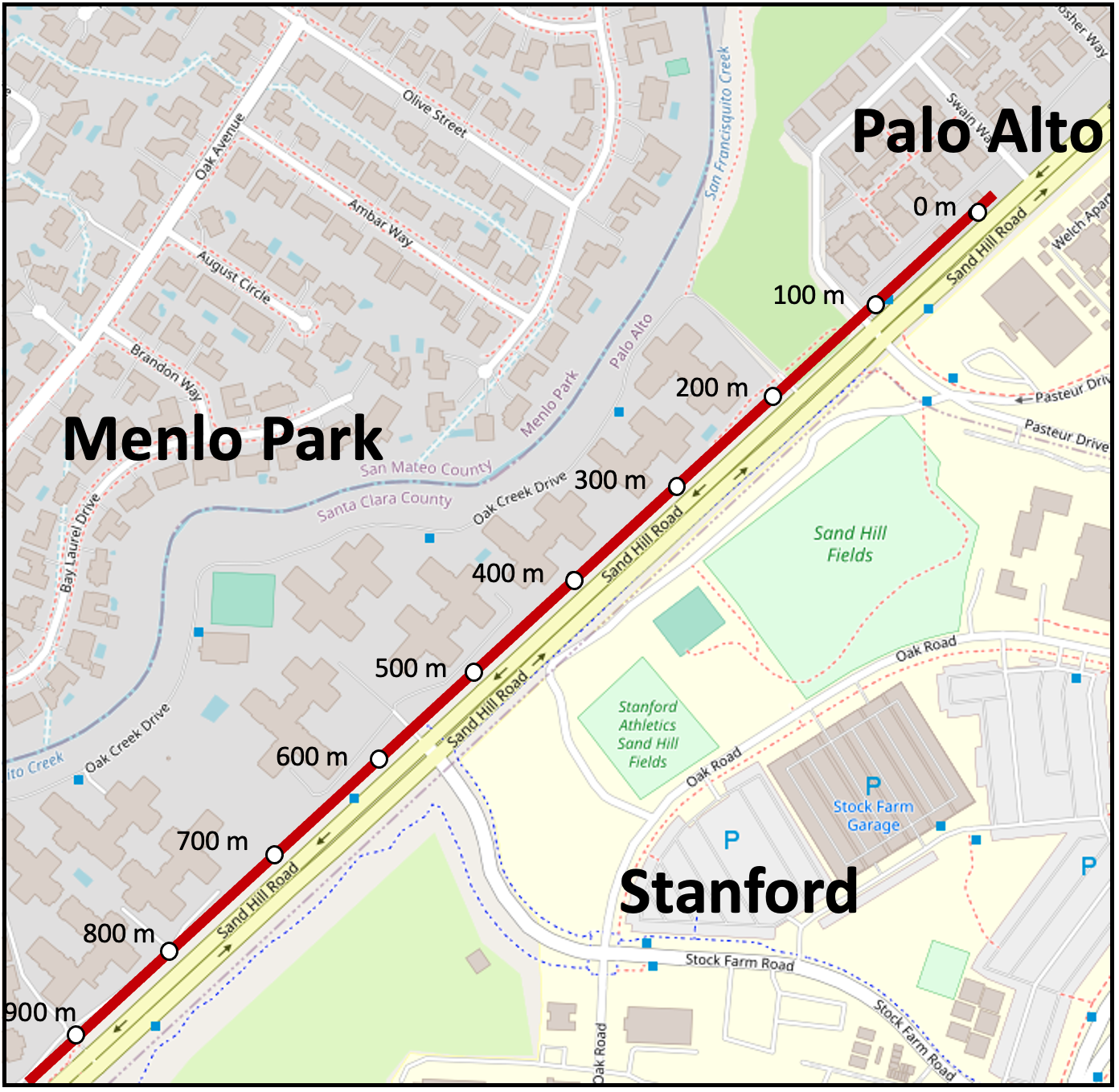}
    \caption{Map view of a roadside section of the Stanford DAS‐2 Array.  Distances along the fiber are labeled on the map}
    \label{fig:map}
\end{figure}

\begin{figure*}[t]
    \centering
    \includegraphics[width=0.7\linewidth]{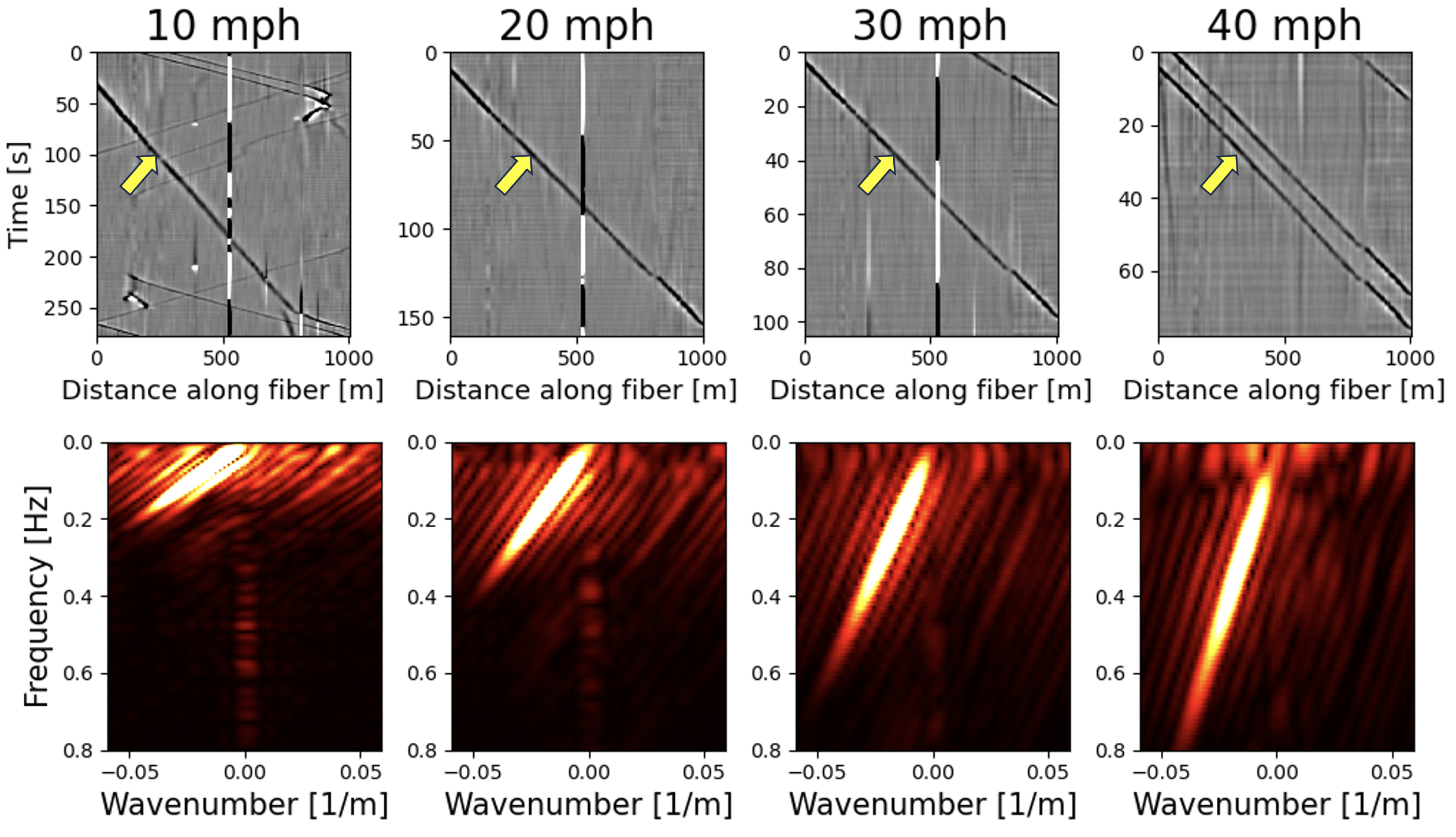}
    \caption{(Top) DAS recordings of the test car driving with constant speeds of 10, 20, 30 and 40 mph. The time axes are scaled differently. (Bottom) The corresponding F-K spectra of the DAS recordings above. We can see that the frequency range becomes broader with increasing speed, whereas the wavenumber components are invariant to car speeds. }
    \label{fig:field_experiments}
\end{figure*}

\noindent Our observation that the shape of the space-domain rather than the time-domain wavelet is speed-invariant is confirmed using field experiments at Sand Hill road monitoring by the Stanford DAS2 array \cite{YuanEtAl2020}. Figure~\ref{fig:map} shows the map of a subsection of the roadside fiber labeled with distances along the fiber. The horizontal distances from the centers of the south- and north-bound traffic lanes to the fiber are around 15 m and 25 m, respectively. We drove a test car southward to understand the car's impulse responses for various driving speeds, including 10, 20, 30, and 40 mph. Figure~\ref{fig:field_experiments} shows the quasi-static signals of our car indicated from DAS recordings using orange arrows. The bottom panels show the corresponding Frequency-Wavenumber spectra. It can be seen that the frequency components vary with speeds, i.e., a higher car speed leads to a broader bandwidth (narrower temporal wavelet). We can also observe that the wavenumber components remain relatively invariant to car speeds, implying that the spatial wavelet for different speeds is stationary.    

\section{Methods}

The previous section brought attention to a key issue: the degradation of resolution in quasi-static vehicle signals, primarily due to the substantial smoothing effect induced by large fiber-road offset and gauge length. To combat this, we propose a spatial-domain DAE model that enhances resolution through spatial-axis deconvolution.

Our advanced spatial-domain DAE model evolves from the original time-domain DAE model that introduced in \cite{van2021deep}. The time-domain model proves effective under conditions of roughly constant vehicle speed over time and when the temporal wavelet aligns with their proposed Ricker wavelet. However, as underscored in the previous section, the speed-dependent variation in the vehicle's time-domain impulse response can cause wavelet mismatch in the time domain, leading to suboptimal deconvolution outcomes, a topic further explored in Section V.

This section begins with an outline of the network architecture, followed by a detailed explanation of our novel extension to the original time-domain model. As will be elaborated in Section V, our innovative extension enables a more resilient model, adept at generating high-quality outcomes that are unaffected by speed variations.

\subsection{Network Architecture} 
\begin{figure*}[t]
    \centering
    \includegraphics[width=0.9\linewidth]{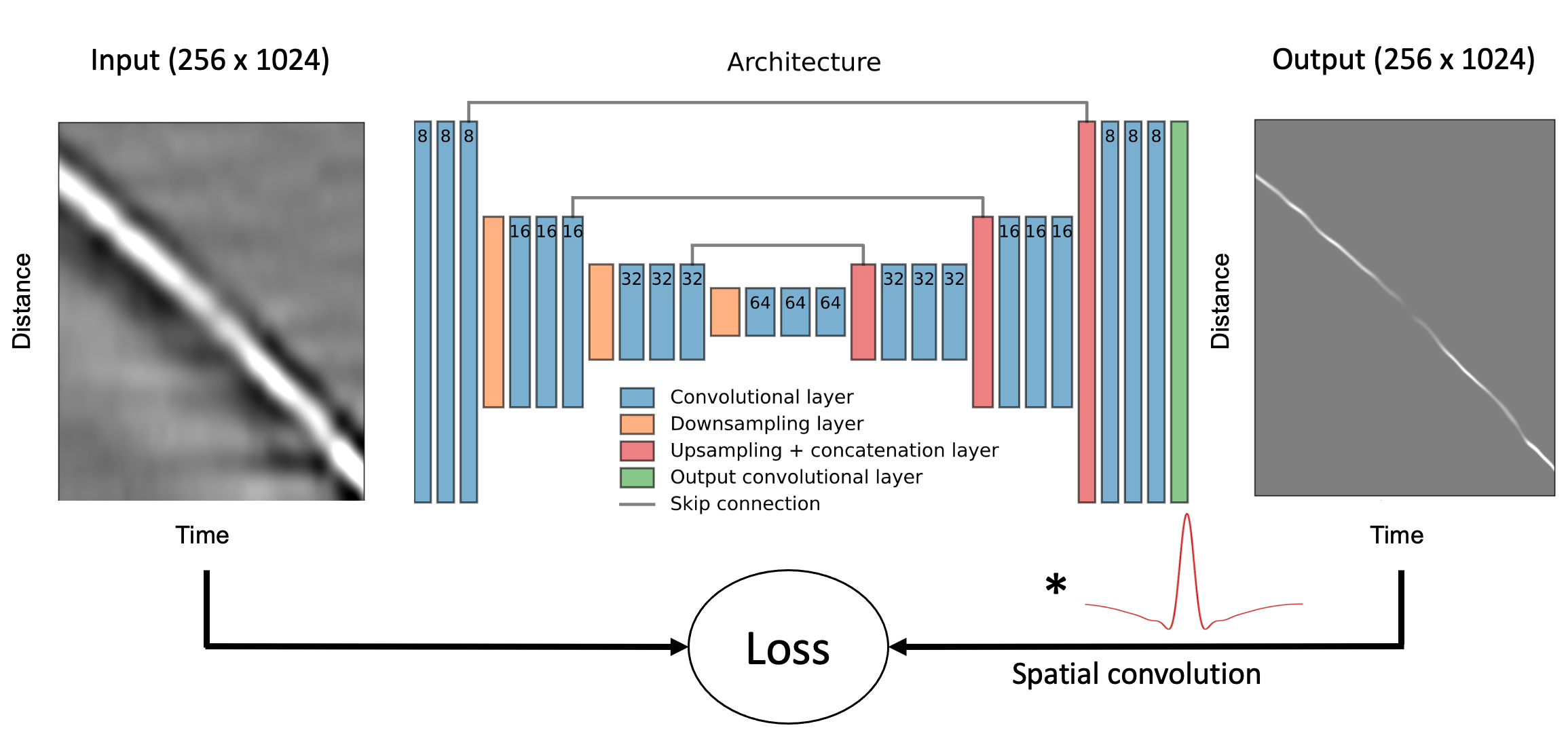}
    \caption{Conceptual overview of the spatial DAE model. The input is the quasi-static response of DAS to cars, which can be viewed as a matrix, 256 channels (256 meters) x 1024 time steps (20.48 seconds). The output is the deconvolution results. The loss is computed with the input and the reconstructed input is obtained through a spatial convolution of the output with a known car impulse response in the spatial domain. }
    \label{fig:U_net_model}
\end{figure*}

Our space-domain DAE model shown in Figure~\ref{fig:U_net_model} is a 2-D fully convolutional U-Net adapted from \cite{van2021deep}. The inputs are quasi-static traffic recordings, a set of $N_x=256$ (256 meters) consecutive waveforms of $N_t = 1024$ time samples (20.48 s) in duration, organized in an $N_x \times N_t$ matrix. The outputs are sharp deconvolution results with the same shape as the inputs. The U-Net model comprises 3 convolutional layers, followed by 3 encoder blocks containing a downsampling (max pooling) layer and 3 convolutional layers. The kernel sizes for the convolution layers are 3 × 5. The number of convolutional filters is initialized at 8 and gets doubled after each downsampling operation. The maxpooling operation downsamples the data by a factor of 2 along the DAS sensor axis and by a factor of 4 along the time axis (i.e, the maxpooling kernel and strides are of size 2 × 4). The decoder reverses the encoding operations with 3 blocks of bilinear upsampling. The U-Net contains skip-connections, which directly connect the output of one encoder block with the corresponding decoder block. Lastly, the output layer is a single convolutional layer with 1 output channel and ReLU activation, which enforces positivity in the model output. Our model is a semi-supervised algorithm, in the sense that no ground truth deconvolution is required as labels to train the model. Weak supervision comes from the spatial car impulse response kernel shown as the red curve in Figure~\ref{fig:U_net_model}.

\subsection{Distinguishing Features of Our space-domain DAE Model}
The primary distinction between our DAE model and the conventional time-domain DAE model in \cite{van2021deep} lies in the fact that we use a simulated wavelet in the spatial domain as opposed to a Ricker wavelet in the time domain. A principal advantage of this approach is that the impulse response in the space domain is not affected by speed and can be associated with the number of axles and the wheelbase. As a result, our innovative application of a spatial-domain wavelet allows us to yield reliable results for vehicles with varying speeds, and retrieve the number of axles, and estimate the wheelbase for larger vehicles.

Additionally, our simulated wavelet is estimated based on the physics defined in Equations \eqref{ux} through \eqref{uxt}. This grounding in physics facilitates the integration of the gauge length and lateral fiber-road offset smoothing effects, which results in a more accurate wavelet and subsequently enhances the performance of our model.

When dealing with a batch of quasi-static inputs, ${y_1, y_2, \dots, y_{N_b}}$ ($N_b$ being the batch size, $N_b=128$), the loss function is formulated as a combination of the L-2 norm of the difference between the reconstructed input and the original input. Because of the use of the spatial wavelet, our reconstructed input is obtained with spatial convolution between the space-domain impulse response and network output. This deviates from the temporal convolution in \cite{van2021deep}. In contrast to the exclusive use of the L-1 norm regularization term in previous work, we explore both the L-1 and L-2 norm of the outputs. This extension allows us to determine optimal settings to more accurately extract the weaker signals of vehicles traveling in distant lanes. Our loss function is defined as follows:
\begin{equation}
\label{obj_dae}
\mathcal{L} = \frac{1}{N_b}\sum_{i=1}^{N_b}(||[k*x_i]_d - y_i||^2_2 +\rho ||x_i||_{1, 2}),
\end{equation}
where $x_i$ and $y_i$ denote the $i$-th deconvolution output and quasi-static input of the U-Net model, respectively. $\rho$ is a weighting term that promotes sparseness in the deconvolved results. $[*]_d$ refers to convolution along the sensor axis. $k$ stands for the spatial wavelet. As we will see in the following sections, unlike conventional channel-independent linear filters, the 2-D deconvolution operations incorporate spatial-temporal features in the DAS recordings. The non-linear nature of the U-Net introduces high frequencies that are not present in the inputs, producing sharp and localized outputs. 

\subsection{Spatial kernel estimation}
In practice, the spatial kernel can be estimated either through numerical simulation described in the previous section or by performing statistical averaging of responses of multiple passing cars assuming that the spatial impulse response is constant in time at each fiber location. The statistical averaging approach requires detecting several passing cars in a subsection of the fiber. The detection can be achieved through manual inspection. Herein, we apply a find-local-maximum algorithm from the SciPy library \cite{2020SciPy-NMeth} to the recordings at quiet midnight to detect an average waveform from isolated cars.

\section{Training and evaluation baselines}
This section aims to outline the steps involved in training our space-domain DAE models and introducing the conventional methods used for evaluation.

\subsection{Training procedure}
We trained our proposed space-domain DAE models with the simulated spatial wavelets of a point load at $y=15$ and $y=25$ m (targeting traffic at the south- and north-bound lanes, respectively) with a DAS gauge length of 16 m shown as the orange and green curves in Fig \ref{fig:point_sim} (c). The choice of $y$ and gauge length match the real settings of the Sand Hill Road DAS experiments. The training dataset we used is 2-hours' worth of traffic recordings of the Sand Hill DAS fiber. We split the 2-hour recordings into a training and evaluation set with a ratio of $80\%$ to $20\%$. 

\subsection{Evaluation baselines}
As a baseline, we trained a time-domain DAE model with the same dataset. The temporal wavelet used for training is simulated with a car speed of 30 mph (shown in Fig.~\ref{fig:wavelet_comparisons}) matching the speed limit of the Sand Hill road.

We also benchmark our spatial DAE model using a conventional spatial deconvolution algorithm with an objective function:
\begin{equation}
\label{objective_deconv}
\hat{x}_q=\argmin_{x_q}\{\frac{1}{2}||[k*x_q]_d-y_q||_2^2+\rho||x_q||_1\}
\end{equation}
Note that $[k*x_q]_d$ stands for convolution in space between a known spatial impulse response kernel, $k$, and the underlying impulse model, $\hat{x}_q$ stands for the deconvolution output. $y_q$ represents the input. $k$ and $\rho$ have the same meaning as in equation~\eqref{obj_dae}.  One commonly used algorithm to solve this optimization problem is the Iterative Shrinkage Thresholding Algorithm (ISTA; \cite{Chambolle1998, Daubechies2004}). For this study, we adopt an accelerated version of ISTA (Fast-ISTA or FISTA) due to the reason described in \cite{Beck2009}, which exhibits faster convergence guarantees. With FISTA performing spatial deconvolution, signals at each time index are processed independently.

\section{Results}
We exhibit the evaluation outcomes of our proposed space-domain DAE model across a range of applications, including enhancing the precision of car tracking, improving the resolution for vehicles traveling closely, and providing accurate axle count and wheelbase estimates for large vehicles. Additionally, a potential limitation of our approach is discussed, along with a proposed solution to overcome it.

\subsection{Car tracking}

\begin{figure}[t]
    \centering
    \includegraphics[width=1\linewidth]{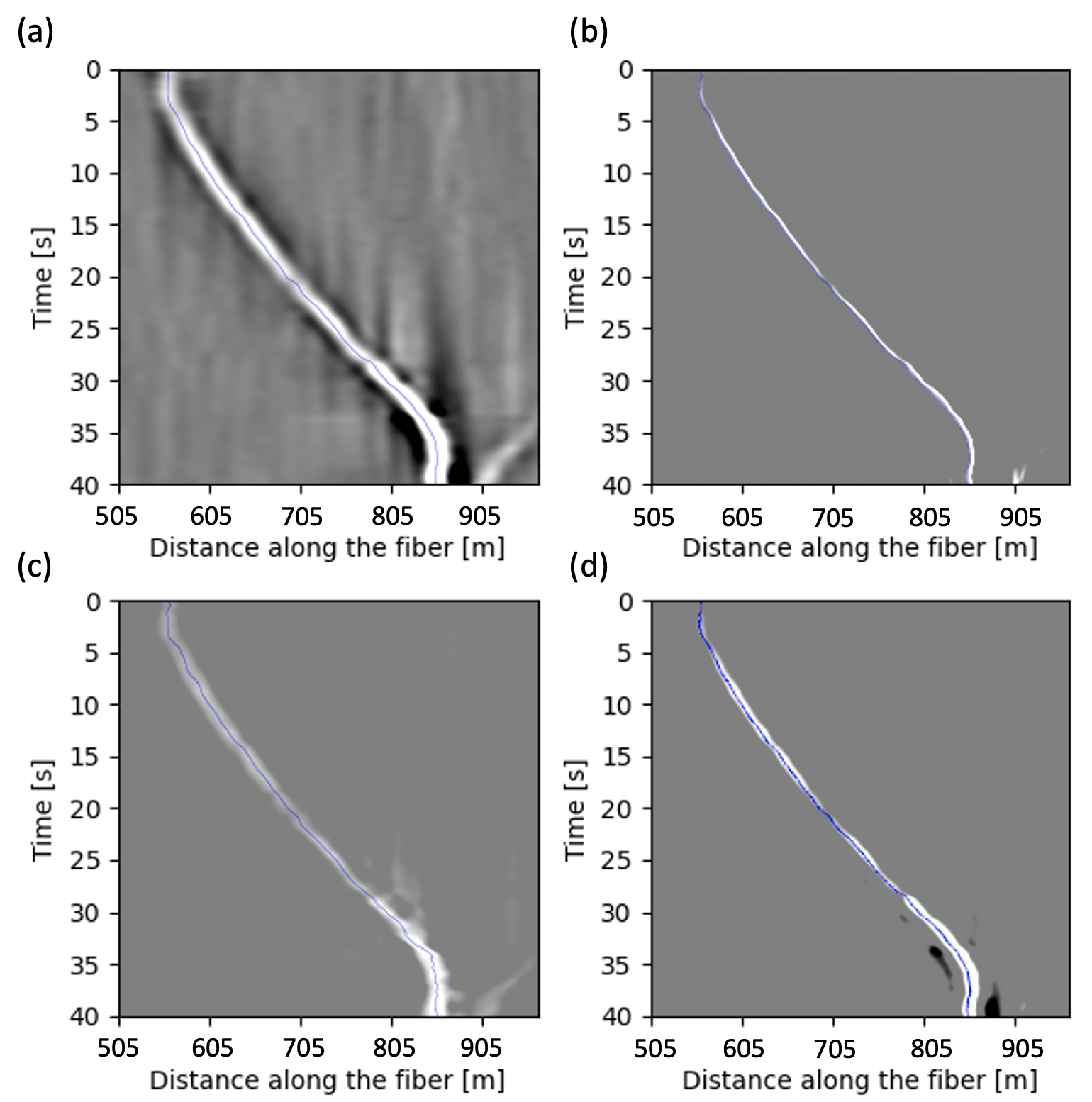}
    \caption{(a) DAS recording of the quasi-static signal of a controlled driving experiment where we first speed up and then slow down; Deconvolution results: (b) the proposed space-domain DAE model, (c) the time-domain DAE model; (d) spatial deconvolution via the FISTA algorithm. The amplitude picks in each panel are shown as a light blue curve reflecting the car trajectory.}
    \label{fig:deconv_comp_t_vs_x_vs_FISTA}
\end{figure}

To test and benchmark the performance of the proposed spatial DAE model, we conducted controlled driving experiments under a rare traffic condition where we drove a test car equipped with a speed sensor and a GPS receiver southward along the subsection of the Sand Hill road shown in Fig. \ref{fig:map}. In Fig.~\ref{fig:deconv_comp_t_vs_x_vs_FISTA}, we focus on a case where our car is first speeding up and then slowing down. Fig.~\ref{fig:deconv_comp_t_vs_x_vs_FISTA} (a) shows the quasi-static signal of our car. The car speed is inversely proportional to the slope in the time-space coordinate. We can see in (a) that when the car speed is low ($< 10$ s and $> 30$ s), the wavelet is ``stretched" in time. When the car speed is relatively higher, the time-domain wavelet is ``compressed", agreeing with our observation in Section II. Fig.~\ref{fig:deconv_comp_t_vs_x_vs_FISTA} (b), (c), and (d) show the deconvolution results from the proposed space-domain DAE model, the original time-domain DAE model, and the space-domain FISTA algorithm. Both the space- and time-domain DAE models are trained with L-1 norm regularization loss to promote sparsity in the outputs. We can see that the proposed space-domain DAE model yields the sharpest and the most localized results regardless of car-speed variation. Meanwhile, we can see that the background noise is suppressed by the DAE model. In contrast, the time-domain DAE model yields results that are dependent on the speed, e.g. the output is more compact in space with a higher speed but becomes stretched out when the speed is lower. The spatial deconvolution via the FISTA algorithm yields results that are speed-invariant but are not as compact as the spatial-DAE results. Signals $< 5$ seconds are oversuppressed by the L-1 norm term in the equation \ref{objective_deconv}. We can also observe artifacts shown in black in the FISTA outputs.

\begin{figure}[t]
    \centering
    \includegraphics[width=1\linewidth]{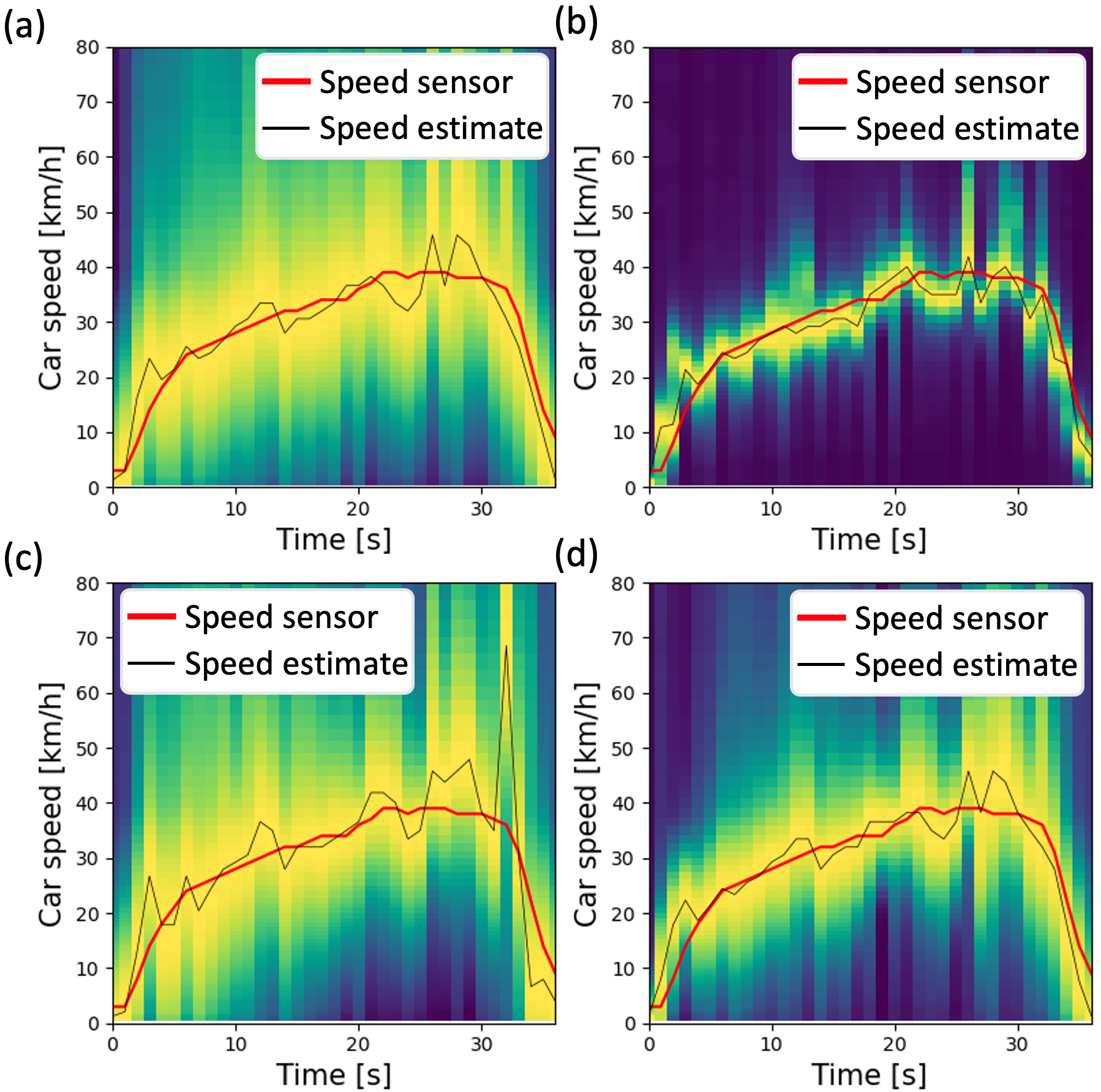}
    \caption{Local beamforming spectra are performed following the trajectory picks in Fig. \ref{fig:deconv_comp_t_vs_x_vs_FISTA} using (a) the quasi-static signal, (b) space-domain DAE results, (c) time-domain DAE results, and (d) FISTA results. The red curve in each panel shows the speed measurements of an onboard speed sensor. The black curves indicate the speed estimates picked from the amplitude of the spectra at each time step.}
    \label{fig:beamforming_comparisons}
\end{figure}

Using the signals in Fig.~\ref{fig:deconv_comp_t_vs_x_vs_FISTA} (a) to (d), we can track the car movement by picking the amplitude at each time step. The estimated trajectory for each panel is shown as a thin blue curve in Fig.~\ref{fig:deconv_comp_t_vs_x_vs_FISTA}. Car speed estimates can be obtained efficiently through a local beamforming algorithm applied to either the quasi-static signals or the deconvolution results (Figure~\ref{fig:beamforming_comparisons}): The beamforming spectrum at each time step is computed using a local 2-D window (3 s x 30 m) following the estimated trajectories. In Fig.~\ref{fig:beamforming_comparisons}, a brighter color indicates higher stack energy in each panel. The red curves represent speed measurements from an onboard speed sensor. Our speed estimates, indicated with the black curves, are obtained by picking the maximum amplitude of the beamforming spectrum at each time step. We can see that our beamforming estimates using the space-domain DAE output match the red curve the best. Using the  Controller Area Network (CAN) bus reading from the vehicle as the ground truth, the root-mean-squared errors (RMSE) for the original recording, space-domain DAE, time-domain DAE, and FISTA are 2.56, 2.14, 4.55, and 2.48 mph, respectively. Our space-domain DAE model achieves the lowest RMSE thanks to the sharp deconvolution outputs and its highest noise suppression performance. The large error of the time-domain DAE model is mainly due to the falsely high-speed event around 33 s as shown in Fig.\ref{fig:deconv_comp_t_vs_x_vs_FISTA} (c) and the wiggled trajectory pickings when the car speed is low ($t>$33~s). 

The selection of a rare traffic scenario for our car tracking case study was intended to accurately assess our method's ability to extract coherent signals from cars varying in speed. Although sparse traffic was used for this evaluation, our system is generally applicable to any traffic conditions.

\subsection{Traffic pattern monitoring}
\begin{figure*}[p]
    \centering
    \includegraphics[width=1\linewidth]{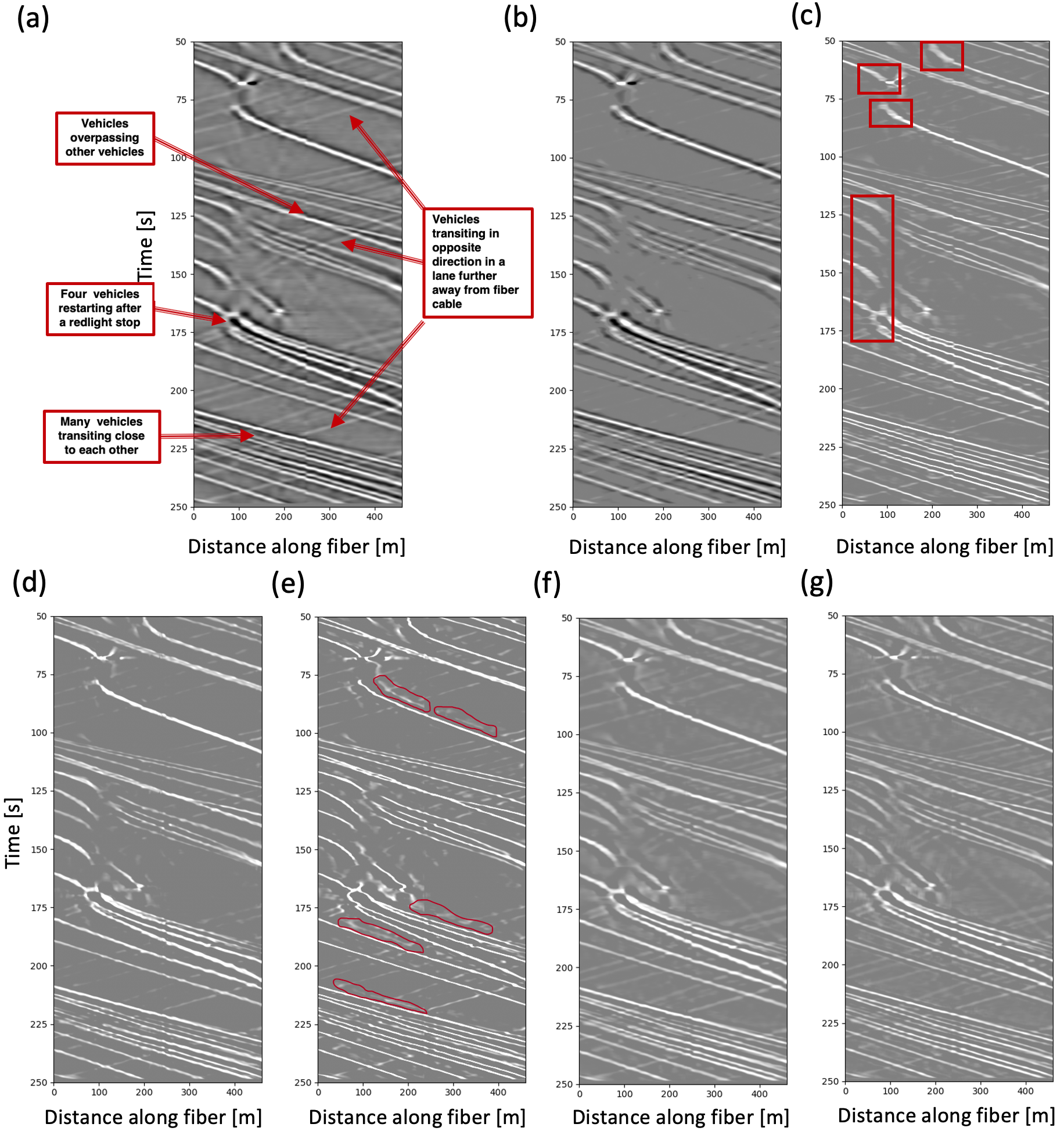}
    \caption{(a) Quasi-static signals of passing vehicles in heavy traffic. (b) spatial deconvolution via FISTA. (c) temporal deconvolution via the time-domain DAE model. Boxes indicate signals in lower resolution due to slower car speeds. (d) and (e) show the results from the proposed space-domain DAE model with L-1 norm regularization using near and far lane wavelets, respectively. Artifacts that could be misinterpreted as transiting cars are circled out. (f) and (g) show the results of the space-domain DAE model with L2 regularization using the near and far wavelets, respectively.}
    \label{fig:heavy_traffic_results}
\end{figure*}

This section investigates the performance of the space-domain DAE model under dense traffic conditions, which we intentionally selected to test the robustness of our model under the most challenging circumstances with numerous closely traveling vehicles.
 Fig.~\ref{fig:heavy_traffic_results} (a) shows a 200-second quasi-static recording in heavy traffic. Vehicles transiting southward are closer to the fiber and thus generate stronger quasi-static signals than the northbound traffic. In the plot, we label several interesting events. We can see that when multiple cars trail closely, their quasi-static signals interfere with each other causing complicated patterns that pose ambiguity for individual car identification. Fig.~\ref{fig:heavy_traffic_results} (b) and (c) show the FISTA and time-domain DAE (L-1 norm regularization) results, respectively. We can see that FISTA generates poorly resolved results and contains evident sidelobe artifacts shown in black. The time-domain DAE model yields sharp results with fast car speed. However, as can be seen from the boxes in Fig.~\ref{fig:heavy_traffic_results} (c), the results get blurred when the car speed is low, e.g. cars stopping for a red light and restarting. Fig.~\ref{fig:heavy_traffic_results} (d)
 shows the results of the proposed space-domain DAE model trained with the near-lane wavelet ($y=15$ m) and L-1 norm regularization loss. We can see that the model yields sharp results regardless of speed changes. However, signals of the northbound traffic in a further lane are oversuppressed. To recover the further lane traffic, we experiment with a spatial DAE model trained with the wider far-lane wavelet (simulated with $y=25$ m matching the fiber-northbound lane offset). Fig.~\ref{fig:heavy_traffic_results} (e) shows the far-lane DAE results. We observe fewer oversuppression issues for the northbound traffic and sharper signals for both near and far lanes. Nonetheless, artifacts that could be misinterpreted as vehicles appear as circled out in the plot. In Fig.~\ref{fig:heavy_traffic_results} (f), we experiment with space-domain DAE models trained with L-2 norm regularization loss using the near-lane wavelet. We can see that with L-2 norm regularization, weak signals of the far-lane traffic are better preserved, which benefits the tracking of the northbound vehicles. Fig.~\ref{fig:heavy_traffic_results} (g) shows the results of the L-2 norm space-domain DAE model trained with the far-lane wavelet. Although the signals are slightly sharper than that shown in (f), the background noise level gets increased. 

\subsection{Large-size vehicle monitoring}
\begin{figure*}[t]
    \centering
    \includegraphics[width=1\linewidth]{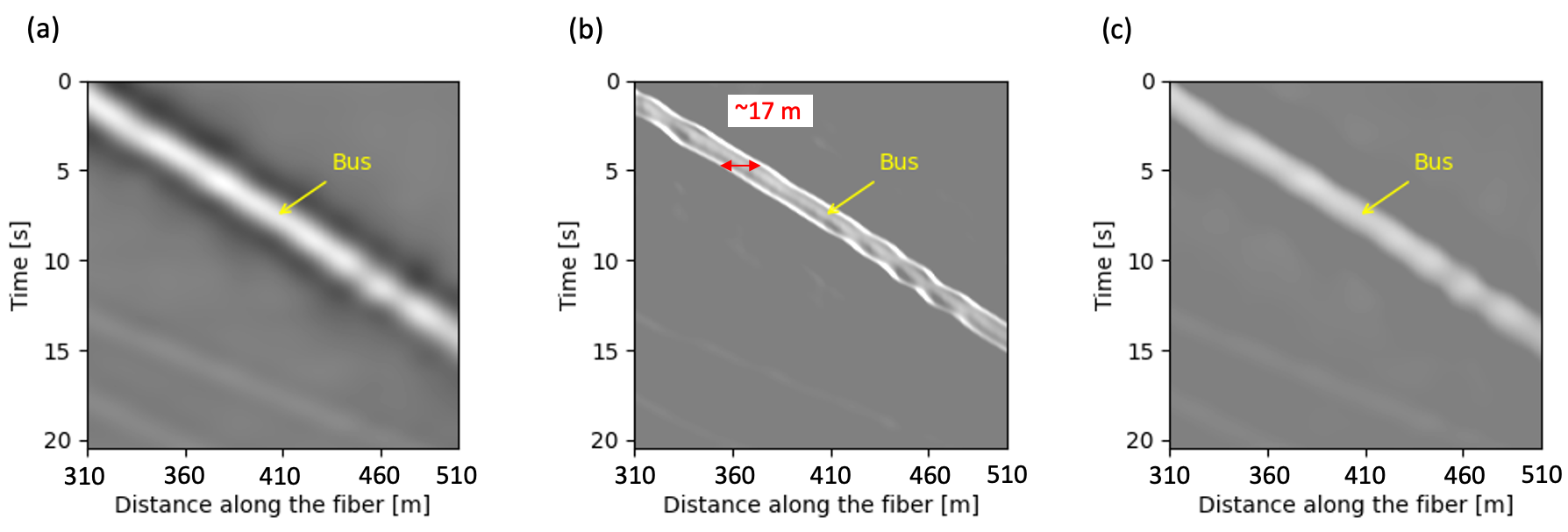}
    \caption{(a) Signals of an 18-meter bus and two regular-sized cars. Deconvolution results of (b) space-domain DAE model with L-1 regularization, and (c) space-domain DAE model with L-2 regularization, respectively.}
    \label{fig:bus}
\end{figure*}

Heavier and longer vehicles, e.g. buses, trucks, and trains, generate quasi-static signals with much larger amplitudes and wider spatial wavelets than cars, e.g., sedans and SUVs. The wider spatial wavelet can be viewed as the superposition of quasi-static signals of wheels at axles that are farther apart. Figure~\ref{fig:bus} (a) shows a quasi-static recording of an 18-meter three-axle bus transiting southward (video captured by our camera). Due to the smoothing effect of the road-fiber offset and gauge length, the axles are indistinguishable from the quasi-static signals. Figure~\ref{fig:bus} (b) shows the deconvolution results of the space-domain DAE model trained with L-1 norm regularization loss using the simulated near-lane impulse response wavelet. We can see the three axles are recovered as three strong-energy trajectories corresponding to the three axles. The amplitude difference between the three trajectories could imply a non-uniform weight distribution on different wheel groups. The distance between the two most substantial peaks is about 17 m agreeing with the bus length, indicating the usage of our L-1 norm model for counting long-size vehicle axles and length characterization. Figure~\ref{fig:bus} (b) shows the results of the space-domain DAE model trained with L-2 norm regularization. We can see that axles are smoothed out by the L-2 norm regularization. The L-2 norm model could be used to track bus motion.

\begin{figure}[t]
    \centering
    \includegraphics[width=1\linewidth]{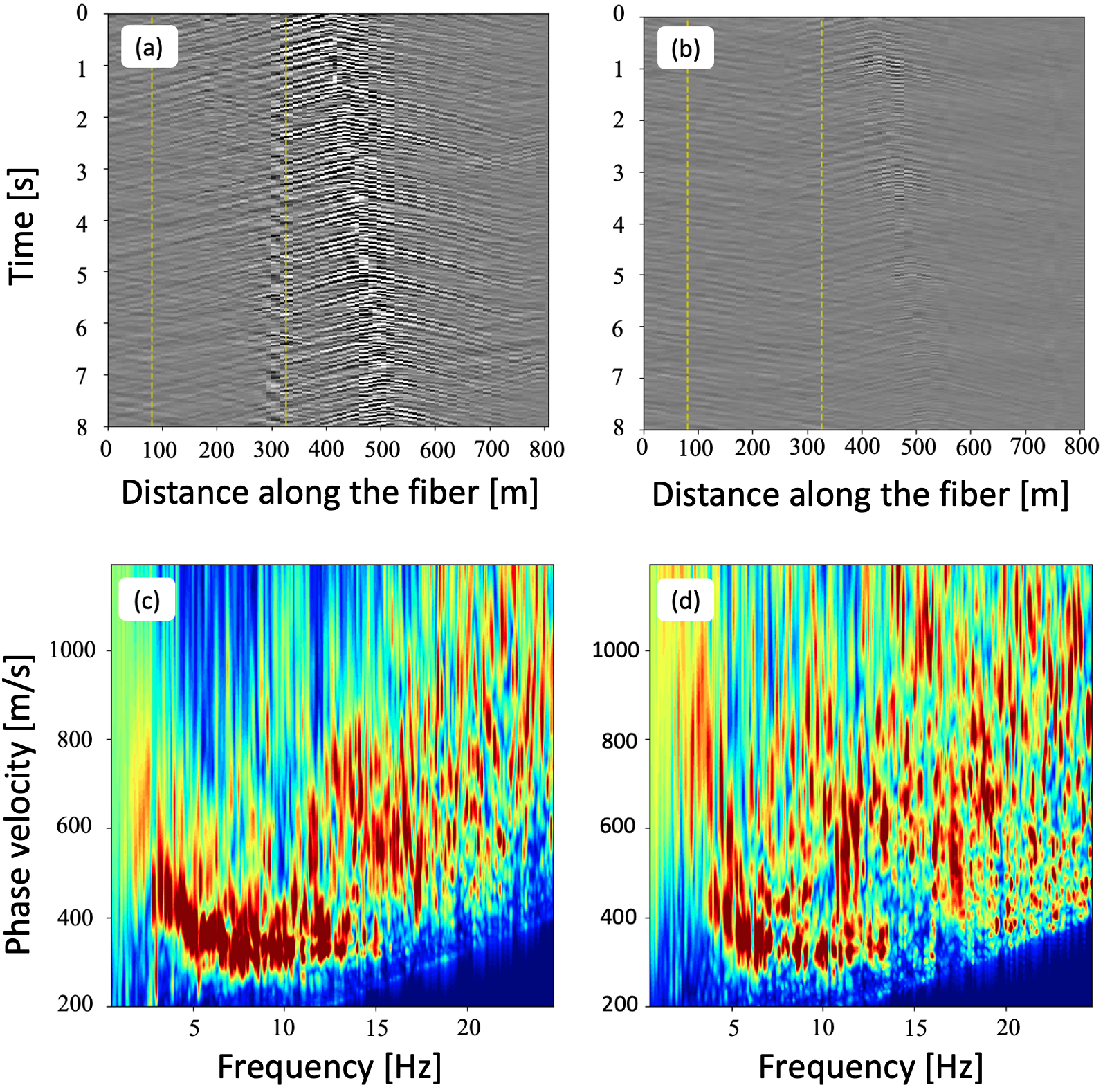}
    \caption{(a) Surface-wave components of (a) a passing bus and (b) a car. (c) and (d) show the dispersion image estimated from the surface waves in (a) and (b), respectively}
    \label{fig:surface_wave}
\end{figure}

\noindent Monitoring large-size vehicles benefits not only traffic management but also non-intrusive shear-wave velocity imaging, which can support city sustainability on applications including sinkhole detection, excavation monitoring, and earthquake hazard analysis. A shear-wave profile can be estimated through an optimization algorithm using traffic-induced surface waves as inputs. Fig.~\ref{fig:surface_wave} (a) and (b) show the dynamic surface wave components ($> 1$ Hz) excited by a moving bus and a car, respectively. We can see that the bus excites much stronger surface waves than the car. With the surface waves within the two dashed lines, we compute the dispersion spectrum reflecting the relationship between the phase velocity and frequency as shown in Fig.~\ref{fig:surface_wave} (c) and (d) for the bus and car, respectively. We can see that large-size vehicles, such as a bus, generates surface waves with frequency as low as $2.5$ Hz, which is absent from the car-induced surface waves. Low frequencies at $2.5$ Hz have a wavelength of $\sim 200$ m \cite{YuanEtAl2020}. The maximum investigation depth of the shear-wave velocity structure using surface waves is about half of the longest wavelength \cite{park2002optimum}. Therefore, with buses, we can retrieve depths down to 100 m, whereas using only the car, results at depths below approximately 50 m are unreliable.

\subsection{Spatial non-stationarity}

\begin{figure}[t]
    \centering
    \includegraphics[width=1\linewidth]{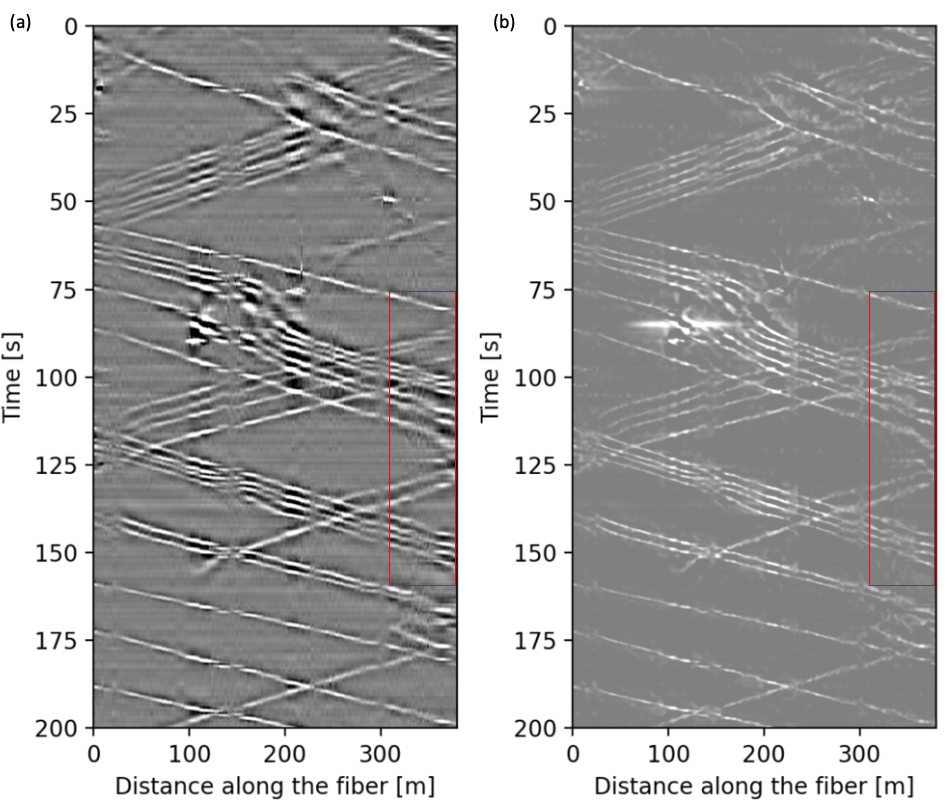}
    \caption{Deconvolution of traffic recording in downtown SJC via a space-domain DAE model with a simulated impulse response stationary in space. (a) Input data; (b) deconvolution results. The red box indicates results that are poorly resolved due to the spatial non-stationarity of the car wavelets.}
    \label{fig:sjc_decon}
\end{figure}

\begin{figure}[t]
    \centering
    \includegraphics[width=1\linewidth]{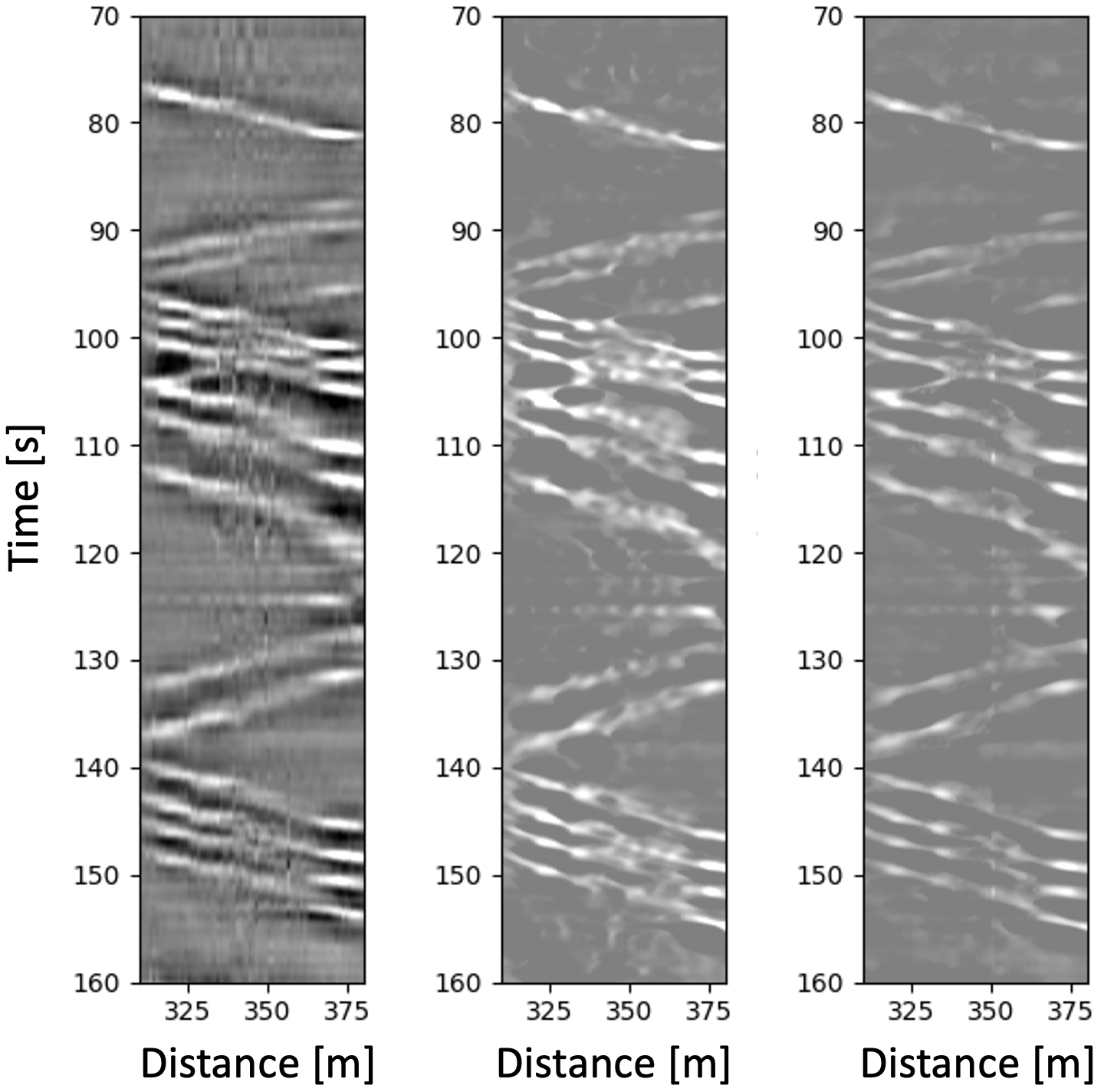}
    \caption{(a) and (b) are respectively zoomed-in views of the data and deconvolution results in the red box shown in Figure~\ref{fig:sjc_decon}. We can see artifacts in (b) due to the wavelet mismatch. (c) shows the better-resolved deconvolution results with a statistical estimate of spatial wavelet for this zoomed-in location.}
    \label{fig:sjc_decon_zoomin}
\end{figure}

Our spatial model assumes that the car impulse response is stationary in space. However, the impulse response is a function of the recording system and the near-surface conditions surrounding the fiber optic cable. Thus, the deconvolution of a DAS array covering heterogeneous near-surface conditions using a single stationary spatial impulse response could be suboptimal. Figure~\ref{fig:sjc_decon} (a) shows DAS recordings of traffic in downtown San Jose City (SJC). (b) shows deconvolution results of the proposed space-domain DAE model with a simulated car impulse response kernel. The deconvolution results look encouraging for most of the fiber, which can be contributed to the consistency of the assumed wavelet to the real wavelet. However, we can see that the results in the red box are relatively poorly resolved. The low resolution is more noticeable from the zoomed-in view of data and deconvolution results in the red box in Figure~\ref{fig:sjc_decon_zoomin} (a) and (b). The under-performance could be explained as the wavelet inconsistency due to the non-stationarity of the near-surface properties. Assuming that the near-surface properties, and in turn the car response, at each channel location is stationary in time. A possible solution would be to estimate the spatial car response at different parts of the fiber by averaging the responses of multiple passing cars around each location. To verify this, we employ a local maximum finding algorithm as a simple car detector for the quasi-static signals. We average the responses of six identified cars passing the area in the red box of Figure~\ref{fig:sjc_decon} as the impulse response input to the U-net model. We retrain the U-net using this estimated kernel, which produces a sharper result in Figure \ref{fig:sjc_decon_zoomin}, indicating the effectiveness of our approach.

\section{Discussion and conclusions}

This paper focuses on the application of traffic monitoring with car-induced quasi-static signals recorded by an urban DAS array. To denoise the data and to reduce the interference among closely traveling cars, we propose a self-supervised convolutional U-Net model (space-domain DAE model) that can compress the quasi-static signals into sharp pulses and remove the background noises. The goal is achieved through spatial deconvolution with an assumed spatial wavelet of the quasi-static signals, which is a major difference from the previously proposed time-domain DAE model. \\

This paper shows that using the spatial instead of the temporal kernel is advantageous because it is invariant to car speed variation. This leads to improved precision and robustness of tracking cars with varying speeds, which is essential for driving behavior identification and accident detection. The benefits of our DAE model are more obvious for heavy-traffic conditions where car signals interfere. With our space-domain DAE model, compressing signals of individual cars to sharp pulses, traffic patterns are much better revealed. Furthermore, we show that our space-domain DAE model is robust to deconvolve signals of large-size and heavy vehicles, such as a bus. Being able to identify large vehicles can be helpful in extracting low-frequency surface waves to image the velocity structure down to hundreds of meters beneath the fiber. Lastly, we point out that a limitation of our spatial-domain DAE model arises when car impulse responses at different locations along the fiber vary due to coupling heterogeneity and/or fiber properties. We address the issue by extracting car wavelets using statistical averaging, and train separated networks. Future work would be adapting the DAE model to different parts of the fiber using location-dependent spatial wavelets. 

\section*{Acknowledgments}
This research was supported financially by the affiliates of the Stanford Exploration Project and the Stanford Sustainability Initiative and the UPS Foundation Endowment Fund. Jingxiao Liu is supported by Leavell Fellowship on Sustainable Built Environment at Stanford University. Martijn van den Ende was supported by the French government through the 3IA Côte d’Azur Investments in the Future project managed by the National Research Agency (ANR) with the reference number ANR-19-P3IA-0002. The interrogator unit was loaned to us by OptaSense Inc. We thank Martin Karrenbach, Victor Yartsev, and Lisa LaFlame from Optasense, as well as the Stanford ITS fiber team, and in particular Erich Snow, for crucial help with the Stanford DAS-2 experiment. We also thank the Stanford School of Earth IT team for hosting the interrogator in the Scholl computer room.


 
\bibliography{bib}
\bibliographystyle{IEEEtran}

%












\newpage


\begin{IEEEbiography}
[{\includegraphics[width=1in,height=1.25in,clip,keepaspectratio]{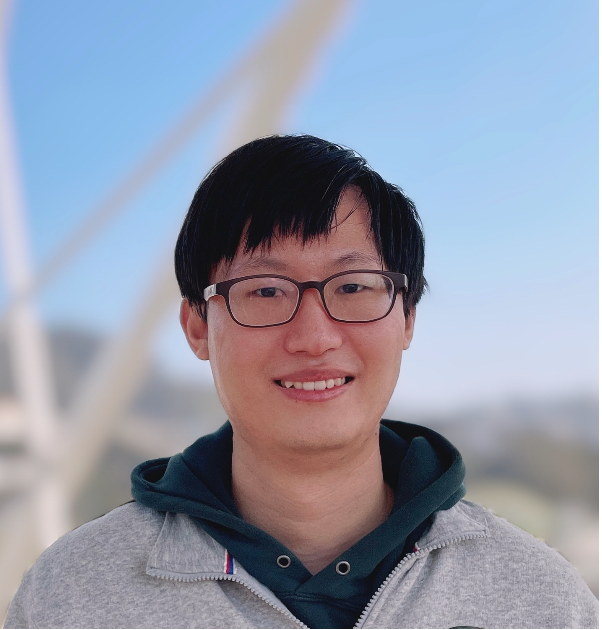}}]{Siyuan Yuan}
is a Ph.D. candidate in Geophysics at Stanford University. His research involves digital city applications of Distributed Acoustic Sensing (DAS) repurposing pre-existing telecommunication fibers as large-scale sensors for urban traffic monitoring, near-surface imaging with vehicle-induced surface waves, car-based fiber mapping, and structural health monitoring. Additionally, he is interested in Machine Learning based speech enhancement technologies. Before his Ph.D. study, he received a M.S. degree in Civil Engineering from Stanford, and a B.S. degree in Civil Engineering from Tongji University, China. 
\end{IEEEbiography}

\begin{IEEEbiography}
[{\includegraphics[width=1in,height=1.25in,clip,keepaspectratio]{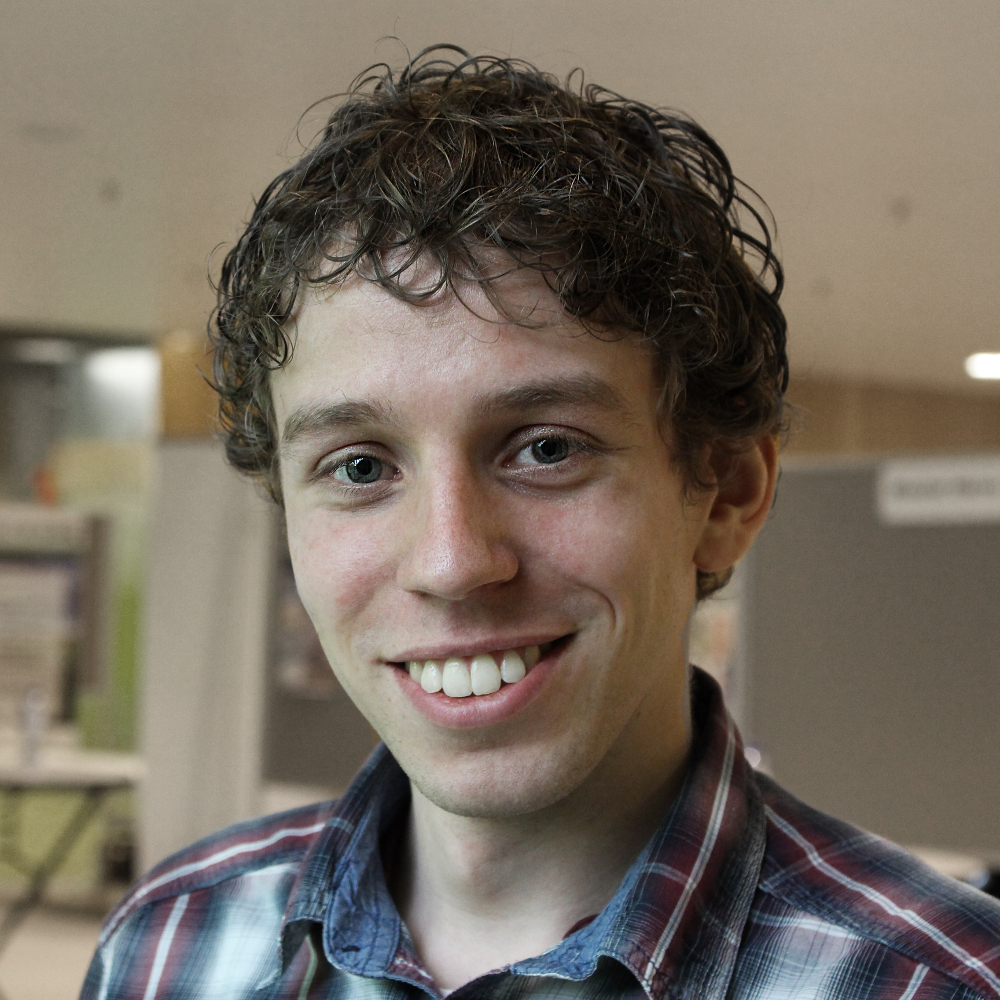}}]{Martijn van den Ende}
received his BSc and MSc degrees in Earth Sciences in 2011 and 2013, respectively, from Utrecht University, the Netherlands. In 2018 he received a PhD degree in fault mechanics from the High Pressure and Temperature laboratory at Utrecht University. Since then, he has joined the Géoazur laboratory (Université Côte d'Azur) to work on earthquake cycle numerical modelling, seismic array processing, and Distributed Acoustic Sensing (DAS), aided by Deep Learning. His latest research involves the identification and analysis of (micro)earthquakes recorded with submarine DAS along the Chilean margin, paving the way for the world's first DAS-based earthquake early warning system.
\end{IEEEbiography}

\begin{IEEEbiography}
[{\includegraphics[width=1in,height=1.25in,clip,keepaspectratio]{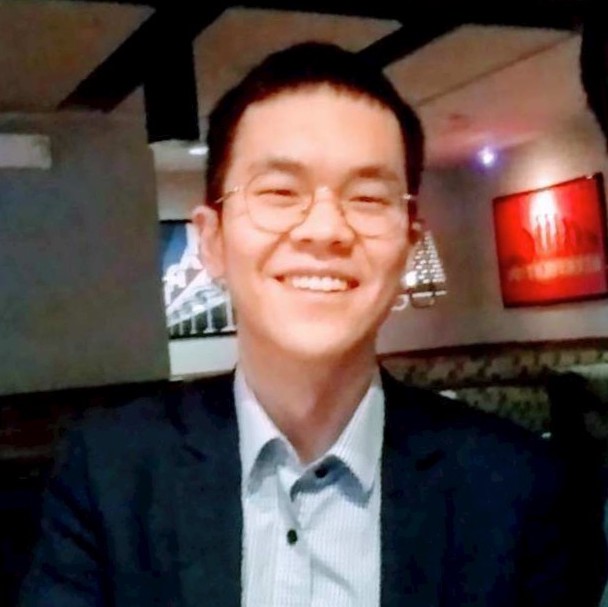}}]{Jingxiao Liu}
is a post-doctoral fellow of Geophysics at Stanford University. He received his Ph.D. in the Department of Civil \& Environmental Engineering with a Ph.D. minor in Electrical Engineering at Stanford University. His research focuses on structural health monitoring, smart infrastructure systems, and smart city applications integrating structural dynamics, signal processing, physics-guided machine learning, mobile sensing, and fiber-optic sensing techniques. He received his M.S. in Civil Engineering from Carnegie Mellon University, and his B.S. in Civil Engineering from Central South University, China. He received the Leavell Fellowship on Sustainable Built Environment and various best paper and presentation awards from ASCE, ASME, and ACM conferences.
\end{IEEEbiography}

\begin{IEEEbiography}
[{\includegraphics[width=1in,height=1.25in,clip,keepaspectratio]{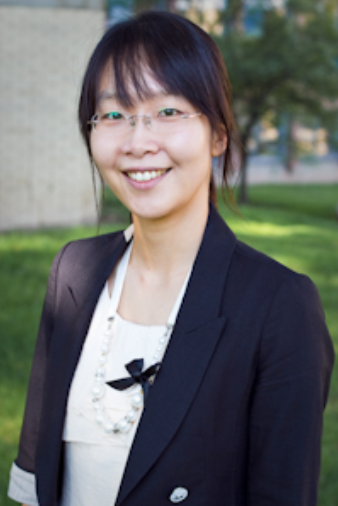}}]{Hae Young Noh}
 is an Associate Professor in the Department of Civil and Environmental Engineering at Stanford University. Her research focuses on indirect sensing and physics-guided data analytics to enable low-cost non-intrusive monitoring of cyber-physical-human systems. She is particularly interested in developing structures to be self-, user-, and surrounding-aware to improve users’ quality of life and provide safe and sustainable built environment. The results of her work have been deployed in a number of real-world applications from trains, to the Amish community, to eldercare centers, to pig farms. Before joining Stanford, she was a faculty member at Carnegie Mellon University. She received her Ph.D. and M.S. degrees in Civil and Environmental Engineering and the second M.S. degree in Electrical Engineering at Stanford University. She earned her B.S. degree in Mechanical and Aerospace Engineering at Cornell University. She received several awards, including the Google Faculty Research Awards (2013, 2016), the Dean’s Early Career Fellowship (2018), the NSF CAREER Award (2017), and various Best Paper Awards from ASCE, ASME, ACM, IEEE, and SEM conferences.
\end{IEEEbiography}

\begin{IEEEbiographynophoto}{Robert Clapp}
grew up in the northwest, spending most of his childhood in Eugene, Oregon. In 1993 he received my B.Sc. (Hons.) in Geophysical Engineering from Colorado School of Mines. He then began graduate studies in the Geophysics Department at Stanford University. He joined the Stanford Exploration Project (SEP) headed by Jon Claerbout and Biondo Biondi. He received his Masters in 1995 and his Phd in 2000. The title of his thesis was Migration velocity analysis with geologic constraints. Since 2000 he has held various titles (PostDoc, Research Associate, and now Senior Research Engineer) within the Geophysics Department. He now works half time with SEP and half the time with the Center for Computational Earth and Environmental Science (CEES). His current research interests include velocity analysis, imaging, computational interpretation, high-performance computing, hardware accelerators, and parallel computing. In 2006 he received the J. Clarence Karcher award from the SEG. He is a member of the SEG and AGU.
\end{IEEEbiographynophoto}

\begin{IEEEbiography}[{\includegraphics[width=1in,height=1.25in,clip,keepaspectratio]{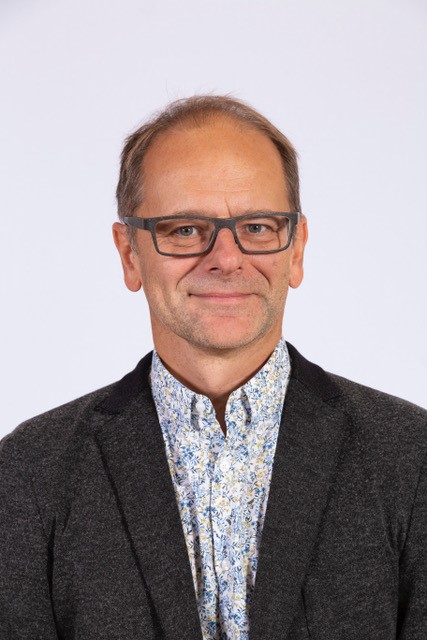}}]{Cédric Richard}
(Senior Member, IEEE) received the Dipl.-Ing. and the M.S. degrees in 1994, and the Ph.D. degree in 1998, from the Compiègne University of Technology, Compiègne, France, all in electrical and computer engineering. He is a Full Professor with Université Côte d’Azur, Nice, France. He is the author of more 300 journal and conference papers. His current research interests include statistical signal processing and machine learning.

Dr. Richard is the Director of the French federative CNRS research association ISIS (Information, Signal, Image, Vision). In 2010–2015, he was distinguished as a Junior Member of the Institut Universitaire de France.

Dr. Richard, since 2020, has served as a Senior Area Chair for the IEEE Signal Processing Letters and, since 2019, as an Associate Editor for the IEEE Open Journal of Signal Processing. In 2015–2018, he was also a Senior Area Chair for the IEEE Transactions on Signal Processing, and an Associate Editor for the IEEE Transactions on Signal and Information Processing over Networks. He was an Associate Editor for the IEEE Transactions on Signal Processing (2006–2010). He is the Vice-Chair of the IEEE Signal Processing Theory and Methods Technical Committee. In 2019–2020, Prof. Richard served as the Director-at-Large of Region 8 (Europe, Middle East, and Africa) of the IEEE Signal Processing Society (IEEE-SPS) and as a Member of the Board of Governors of the IEEE-SPS. 
\end{IEEEbiography}

\begin{IEEEbiography}[{\includegraphics[width=1in,height=1.25in,clip,keepaspectratio]{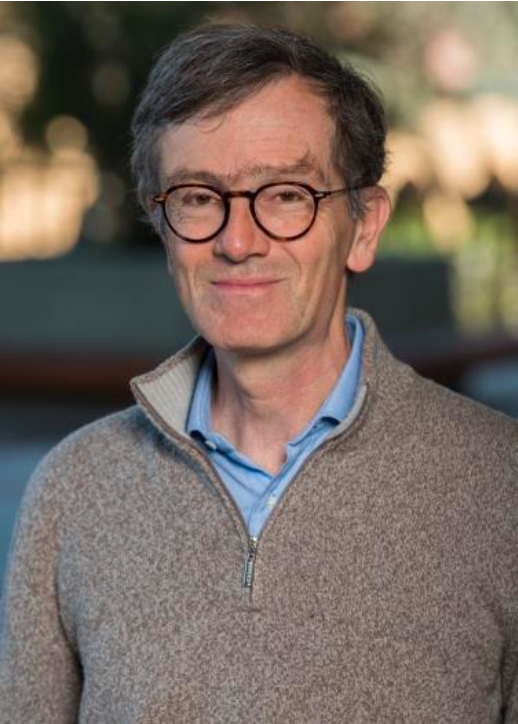}}]{Biondo Biondi} is the Barney and Estelle Morris Professor of Geophysics at Stanford University. He is director of the Stanford Exploration Project (SEP). SEP is an academic consortium whose mission is to develop innovative seismic imaging methodologies and to educate the next generation of leaders in applied seismology.

Biondo and his students devise new algorithms to improve imaging of active and passive seismic data. Images obtained from seismic data are the primary source of information on the structural and stratigraphic complexities in Earth's subsurface and on many subsurface dynamic processes. These images are constructed by processing seismic wavefields recorded at the Earth's surface and generated by either active-source (e.g., vibroseis trucks) experiments or by natural (e.g., ocean waves) and anthropogenic (e.g., vehicle traffic) sources. Because our datasets are enormous, and wavefield propagation needs to be accurately modeled to achieve high-resolution imaging, we need to harness the power of the latest computational hardware to test our methods on field data. Therefore, mapping imaging algorithms into high-performance
architecture is an essential component of our research. The amount and quality of information that we can extract from seismic data are directly linked to the temporal and spatial sampling of the sources and the receivers. In the past several years, we have been working on methods to process data recorded by using fiber cables as seismic sensors. Fiber-optic seismic recording promises to enable cost-effective continuous seismic monitoring at a large scale. A particularly exciting possibility is leveraging preexisting telecommunication infrastructure to record seismic data with dense arrays in urban environments continuously. In 2016 we pioneered that idea by recording data under the Stanford campus. Since then, we recorded data in San Jose and on a 48-km array under Stanford and neighboring cities.
\end{IEEEbiography}

\vspace{11pt}

\vfill

\end{document}